\documentclass[12pt]{article}

\usepackage{scicite}
\usepackage{graphicx}
\usepackage{times}

\newenvironment{sciabstract}{%
\begin{quote} \bf}
{\end{quote}}

\title{Antiepileptic drugs induce subcritical dynamics in human cortical networks} 
\author
{Christian Meisel$^{1\ast}$\\
\\
\normalsize{$^{1}$Department of Neurology, University Clinic Carl Gustav Carus,}\\
\normalsize{Fetscherstra\ss e 74, 01307 Dresden, Germany}\\
\\
\normalsize{$^\ast$To whom correspondence should be addressed; E-mail:  christian@meisel.de.}
}

\date{}

\begin{document} 

% Double-space the manuscript.

\baselineskip24pt

% Make the title.

\maketitle

% Place your abstract within the special {sciabstract} environment.

\begin{sciabstract}

Cortical network functioning critically depends on finely tuned interactions to afford neuronal activity propagation over long distances while avoiding runaway excitation.
This importance is highlighted by the pathological consequences and impaired performance resulting from aberrant network excitability in psychiatric and neurological diseases, such as epilepsy.
Theory and experiment suggest that the control of activity propagation by network interactions can be adequately described by a branching process.
This hypothesis is partially supported by strong evidence for balanced spatiotemporal dynamics observed in the cerebral cortex, however, evidence of a causal relationship between network interactions and cortex activity, as predicted by a branching process, is missing in humans.
Here we test this cause-effect relationship by monitoring cortex activity under systematic pharmacological reduction of cortical network interactions with antiepileptic drugs.
We report that cortical activity cascades, presented by the propagating patterns of epileptic spikes, as well as temporal correlations decline precisely as predicted for a branching process. 
Our results provide the missing link to the branching process theory of cortical network function with implications for understanding the foundations of cortical excitability and its monitoring in conditions like epilepsy.

\end{sciabstract}

% In setting up this template for *Science* papers, we've used both
% the \section* command and the \paragraph* command for topical
% divisions.  Which you use will of course depend on the type of paper
% you're writing.  Review Articles tend to have displayed headings, for
% which \section* is more appropriate; Research Articles, when they have
% formal topical divisions at all, tend to signal them with bold text
% that runs into the paragraph, for which \paragraph* is the right
% choice.  Either way, use the asterisk (*) modifier, as shown, to
% suppress numbering.

\section*{Introduction}

Cortical network functioning critically depends on a finely tuned level of excitability, the transient or steady-state response in which the brain reacts to a stimulus. On the one side, excitability must be small enough to prevent explosive growth of neuronal activity cascades. On the other side, it must be large enough to allow for activity propagation over long distances to afford neuronal communication across sites far apart. 
The importance of finely tuned cortical excitability levels is highlighted by the pathological consequences and impairments resulting from aberrant network excitability in neurological \cite{Badawy2012b} and psychiatric diseases \cite{Masuda2019}. In epilepsy, changes in cortical network excitability are believed to be an important cause underlying the initiation and spread of seizures, i.e. the large non-physiological neuronal activity cascades across time and space \cite{Stafstrom2006,Bazhenov2008,Trevelyan2013}. Pharmacological reduction of excitability consequently constitutes the main treatment approach to control seizures \cite{Bialer2010}.

In the brain, excitability is a product of excitatory and inhibitory network interactions.
To avoid regimes where excitability is too high or too low, these interactions must be finely tuned. 
A growing amount of evidence indicates that this control of activity propagation by network interactions can be adequately described by a branching process \cite{Harris1989,Zaperi1995,Beggs2003,Haldeman2005,Plenz2012,Larremore2013,Meisel2017,Wilting2018}.
In a branching process, activity will remain small and local when interactions are too weak. When interactions are too strong, dynamics over-activates the whole network. At the critical transition between these two states, activity propagates in balanced cascades, or avalanches, avoiding premature die-out and runaway excitation. 
These balanced propagation patterns closely match empirical observations in animal and human studies where spontaneous activity was found to propagate from one active group of neurons to another in cascades over long distances without runaway excitation \cite{Beggs2003,Palva2013,Haimovici2013,Bellay2015}.
Further evidence comes from observations of long-range temporal correlations in cortical activity \cite{Linkenkaer2001,Palva2013,Meisel2017}, another hallmark of a critical branching process \cite{Bak1987,Meisel2017,Wilting2018}.
When network interactions in a branching process are reduced, cascade sizes and temporal correlations decline  \cite{Harris1989,Zaperi1995,Beggs2003,Markram2015,Meisel2017,Wilting2018}.
In vitro studies using cortex preparations, where network interactions can be pharmacologically reduced, show that activity changes closely match the predictions of a branching process \cite{Beggs2003}.

Empirical evidence, however, that alterations in cortical network interactions predict dynamics changes according to a branching process in humans is missing.  
The lack of this cause-effect demonstration constitutes a missing link to the branching process theory with implications for understanding the foundations of cortical excitability and its management in conditions like epilepsy.
Here we directly test the hypothesis that cortical network interactions control dynamics according to a branching process in humans. We make use of the notion that antiepileptic drugs (AEDs) are specifically targeted at reducing network interactions either by reduction of a neuron's individual excitability, reduction of excitatory synaptic transmission or increase in inhibitory synaptic transmission \cite{Bialer2010}. By systematic investigation of the effects of AEDs on cortex dynamics alongside a companion neural network model, we show that changes in network interactions predict spatiotemporal cortex dynamics precisely as expected for a branching process.

%The difficulty to titrate cortical network interactions has made it difficult to test these predictions in humans. 
%A branching process thus provides a promising unified framework to link cortical network interactions to specific activity patterns and network excitability \cite{Beggs2003,Meisel2017,Meisel2015}.
%The foundations of this hypothesis -- that human cortical networks reside in the vicinity of a phase transition controlled by excitability -- remain unstudied experimentally.
%Further, theory and experiment show that a balanced state gives rise to optimal information processing \cite{Haldeman2005,Kinouchi2006,Shew2009}. 
%In the brain, excitability is controlled by the interplay of excitatory and inhibitory (E/I) transmission: when excitation is low and/or inhibition is high, activity will remain small and local. When excitation is too high and/or inhibition too low, network dynamics is at risk of unphysiological bursts of high activity.
%- Wilting paper: second test condition favours exclusion of critical states?

\section*{Results}
%When interactions, or effective connectivity, are low, activity will, statistically speaking, remain small and local. When network interactions are too high,  and vice versa
%These metrics thus capture the well known fact that activity cascades remain, statistically speaking, small and local when interactions are small (subcritical regime), and span the whole network when high (supercritical regime). When network interactions are too high,  and vice versa
%A branching process posits that cascading activity bursts \cite{Zaperi1995} decline in size along with temporal correlations \cite{Linkenkaer2001,Meisel2017,Wilting2018} as network interactions decrease from critical to subcritical.
We first analyzed a parsimonious neuron network model based on a branching process to review how collective cortical dynamics is shaped by network interactions and AED action. Similar models have been used widely to successfully predict the dynamics of tissue from the cortex in humans, monkeys, rats and turtle \cite{Beggs2003,Haldeman2005,Kinouchi2006,Poil2008,Larremore2011,Larremore2013,Shew2011,Shew2015,Meisel2017}.
The model is simple enough to provide insight into the mechanisms governing collective network dynamics yet entails sufficient detail to model relevant aspects of AED action on network interactions. The network consisted of probabilistic integrate-and-fire neurons with all-to-all connectivity, a subset of neurons (20\%) being inhibitory (Fig. \ref{fig_1} B). Our model differed from previous models in that it contains means to mimic AED action to reduce excitability \cite{Bialer2010} (Fig. \ref{fig_1} A): (a) a variable to probabilistically reduce neuron excitability, (b) a scaling parameter by which excitatory synaptic strengths could be downscaled, and (c) a scaling parameter by which inhibitory synaptic strengths could be upscaled.
We studied how the model dynamics in terms of cascading activity and temporal correlations change as a result of decreasing excitability by means of AED action. 

In the absence of AED action, collective dynamics exhibited the well-known phase transition from a quiescent to an active phase when connection strength was increased (Fig. \ref{fig_1} C, black line). Activity propagated in the form of cascades or avalanches \cite{Beggs2003,Kinouchi2006,Larremore2013}. Cascade sizes, quantified by the large cascade fraction, LCF, became larger as interaction strength was increased, whereas temporal correlations, quantified by the autocorrelation half-width, ACW, peaked at criticality \cite{Linkenkaer2001,Meisel2017,Wilting2018} (Fig. \ref{fig_1} C, grey dotted and broken lines, respectively). 
Next, we studied the effect of AED action on these dynamical signatures. LCF and ACW decreased with each AED mechanism of action modelled when dynamics was placed at criticality or in the subcritical regime (Fig. \ref{fig_1} D, E).  
These results illustrate how AED action reduces network interactions and, thereby, provides means for changing the system's control parameter, i.e., network interactions. 
By controlling network interactions, AEDs therefore allow to directly test if network interactions control spatiotemporal cortex dynamics precisely as predicted by a branching process in humans.

\begin{figure}%[htbp]
\centering
\includegraphics[width=0.75\textwidth]{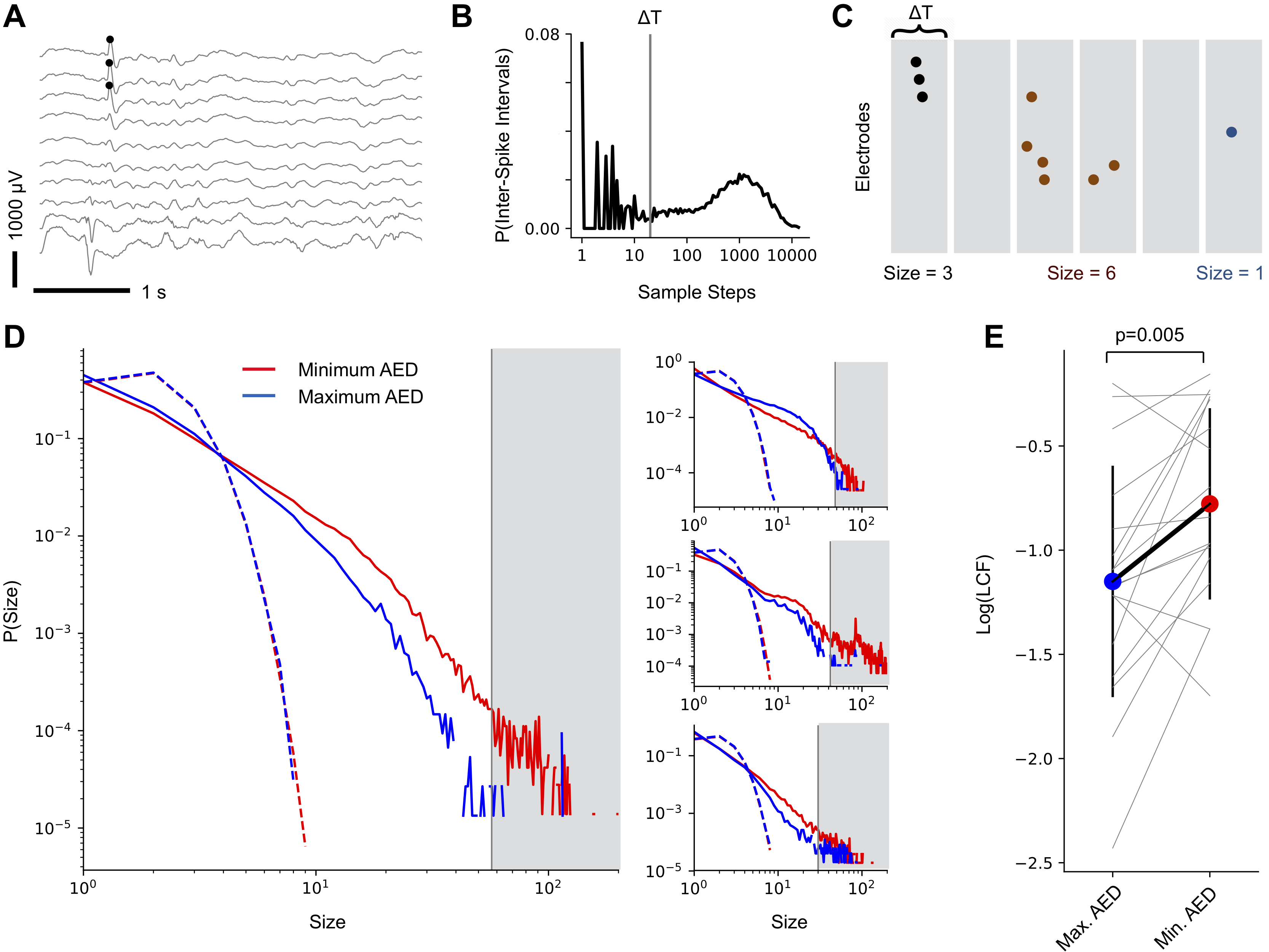}
\caption{Epileptic spikes organize as activity cascades reduced in size by AED action. A, Identification of spikes in electrocorticogram. B, C Bimodality of inter-spike interval distribution identifies spike cascades and their timescale. D, Spike cascade size distribution from four patients under high (blue) and low (red) AED load. High AED load reduces the number of large cascades (shaded grey area). Broken lines indicate size distributions from randomly shuffled spike times. E, AEDs significantly reduce the number of large cascades (large cascade fraction, LCF; grey lines indicate individual patients; black line indicates mean with whiskers denoting standard deviation).\label{fig_2}
}
\end{figure}

To study cascading network events in human cortex, we took advantage of the fact that interictal epileptic spikes superimpose in the extracellular field as a consequence of synchronous activity of spatially neighboured group of neurons (Fig. \ref{fig_2} A). Epileptic spikes consist of elevated population activity known to propagate across cortex \cite{Badier1995}. 
Inter-spike intervals exhibited a bimodal density distribution (Fig. \ref{fig_2} B) indicative of short intervals arising from spikes in the same cascade and long intervals separating different cascades.
Spatiotemporal spike cascades were consequently identified if spikes occurred within the same or consecutive time bins of width $\Delta t$ (Fig. \ref{fig_2} C), where $\Delta t$ was chosen to be greater than the short timescale of inter-spike intervals within a cascade, but less than the longer timescale of inter-cascade quiescent periods \cite{Shew2009}. ​
We observed spikes to organize in cascades of continuous events in time and space indicative of the presence of significant correlations in neuronal activity among cortical sites which, accordingly, were destroyed when the times of spikes were shuffled randomly (Fig. \ref{fig_2} D).
Cascade sizes exhibited a high degree of size variability with larger sizes occurring systematically less often, as predicted by a branching process in the vicinity of criticality or slightly subcritical \cite{Harris1989,Zaperi1995,Beggs2003,Plenz2012,Wilting2018}. Higher antiepileptic drug loads generally led to smaller cascade sizes (Fig. \ref{fig_2} D, blue). As a quantification, the large cascade fraction, LCF, was significantly lower in days with high compared to low AED load (Fig. \ref{fig_2} E, p=0.005, two-sided paired t-test). This effect could not be explained by spike rate changes, which exhibited no difference (p=0.381, two-sided paired t-test).

\begin{figure}%[htbp]
\centering
\includegraphics[width=0.75\textwidth]{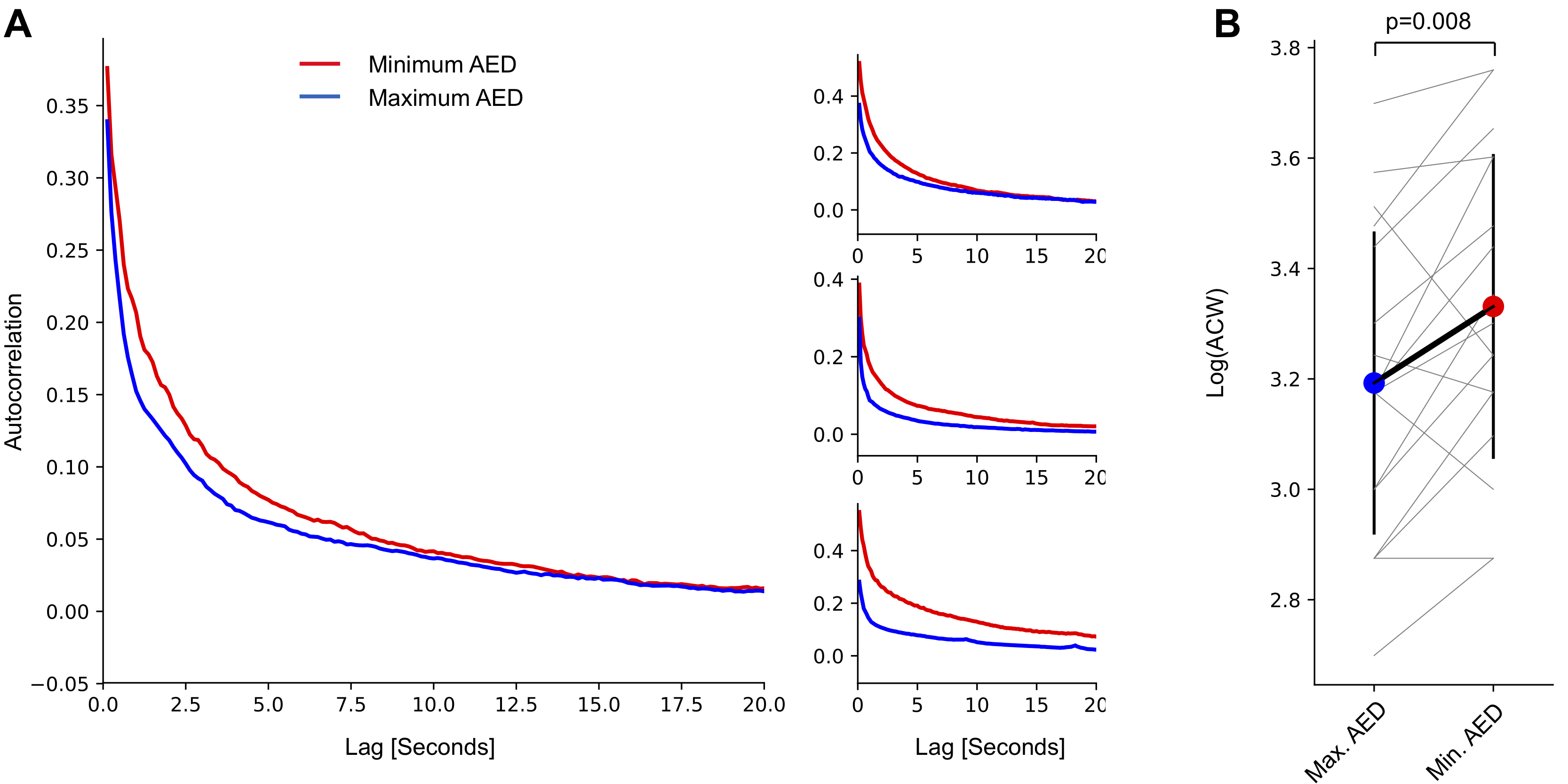}
\caption{AED action reduces temporal correlations in cortex. A, Autocorrelation functions from four patients under high (blue) and low (red) AED load. B, AEDs significantly reduce temporal correlations measured by the autocorrelation function half-width (ACW; grey lines indicate individual patients; black line indicates mean with whiskers denoting standard deviation).\label{fig_3}
}
\end{figure}

Next, we tested whether temporal correlation were controlled by AEDs as predicted by a branching process.
Modulations in signal power are a generally useful currency in characterizing neural dynamics \cite{Donner2011}. We analyzed broadband high-frequency power modulations which provide a local estimate of population spike rate variations near an electrocorticographic electrode \cite{Manning2009, Miller2010, Nir2007, Ray2011, Whittingstall2009}. Autocorrelation functions obtained from these high-frequency power modulation have been shown to accurately capture temporal integration properties \cite{Honey2012}. Autocorrelation functions exhibited a faster decay in high AED medication days compared to low medication days (Fig. \ref{fig_3} A). As a quantification of this decay, the autocorrelation function width, ACW, was significantly lower in days with high compared to low AED load (Fig. \ref{fig_3} B, p=0.008, two-sided paired t-test). This effect could not be explained by changes in high-frequency power, which exhibited no difference (p=0.988, two-sided paired t-test).

\section*{Discussion}
\begin{figure}%[htbp]
\centering
\includegraphics[width=0.3\textwidth]{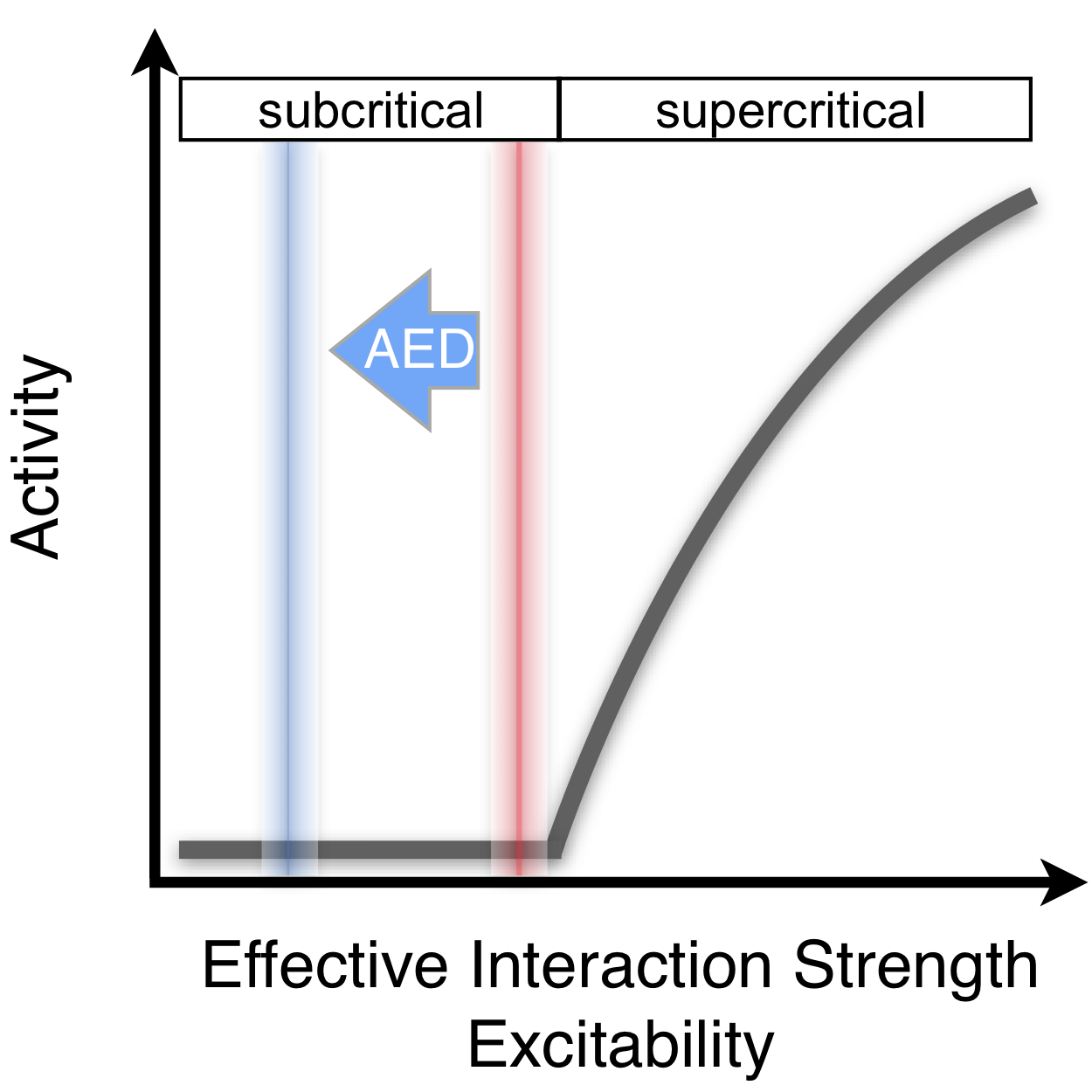}
\caption{Growing evidence suggests that activity propagation in cortical networks can be described by a branching process near the transition between an inactive (subcritical) and an active (supercritical) regime (red). AEDs shift network dynamics further into the subcritical regime (blue), thereby establishing a safety margin to avoid runaway activity associated with the supercritical regime.\label{fig_4}
}
\end{figure}

Our results demonstrate that human cortical dynamics under manipulation of network interactions by AEDs is predicted by a branching process. 
Albeit backed by a large number of computational studies \cite{Kinouchi2006,Poil2008,Larremore2011,Shew2011,Shew2015,Meisel2009,Chialvo2010,Markram2015,Meisel2017}, empirical evidence demonstrating interaction strength as a control parameter in this phase space had previously been limited to reduced in vitro preparations \cite{Beggs2003}.
By using AEDs as means to pharmacologically manipulate and reduce cortical network interactions in epilepsy patients, we report that dynamics becomes more short ranged in terms of spatiotemporal activity cascades and temporal correlations. These findings closely match predictions for dynamical shifts towards the subcritical state, as demonstrated in a companion model. By directly controlling interaction strengths in patients, our work overcomes previous limitations inherent to passive monitoring of network dynamics. Taken together, our results indicate that AEDs drive cortical network dynamics into a subcritical regime by acting on the control parameter which may serve to avoid the risk of runaway excitation (Fig. \ref{fig_4}). 

Beyond providing the missing link to the branching process theory of cortical network function, the current findings have implications for understanding the foundations of cortical excitability and information processing in cortex.
Aberrant excitability levels are an important cause underlying the initiation and spread of seizures \cite{Stafstrom2006,Bazhenov2008,Trevelyan2013}.
Accordingly, the ability to monitor excitability and control its degree is of prime importance for adequate clinical care and treatment. As a unifying framework linking interictal spike cascades, temporal correlations and cortical network excitability, a branching process provides precise markers informed by theory on how to monitor excitability levels from EEG \cite{Meisel2015}. For example, while previous work has shown that interictal spike count itself does not reflect excitability \cite{Stacey2011}, a branching process, backed by our empirical findings, suggests that cascade sizes of interictal spikes are more informative about network excitability. The diminished spread of neural activity to other cortical sites indicated by the smaller cascades under high AED load is in line with observations of lower synchrony under AED \cite{Meisel2015} indicative of decreased cortical interactions. Collectively, a fundamental dynamical understanding of how excitability and its control represents in cortical activity may help to screen and evaluate treatments targeted at excitability in epilepsy and beyond.

The maintenance and integration of information over extended periods of time is considered to be important for information processing at the neural network level \cite{Kiebel2008,Friston2012} for which long-range temporal correlations are thought to provide the neural basis \cite{Honey2012,Chaudhuri2015,Kringelbach2015,Meisel2017}. Consequently, theory and experiment show that a balanced state, where long-range temporal correlation peak, gives rise to optimal information processing \cite{Haldeman2005,Kinouchi2006,Shew2009,Palva2013}. Our results demonstrate a systematic decrease in temporal correlations with AED load, some of which are known have detrimental effects on cognition. The insights gained into the decline of spatiotemporal correlation as a function of decreased cortical excitability may thus help to uncover the underlying neuronal correlates linked to these cognitive impairments.

%\bibliography{sd_slowing}

%\bibliographystyle{Science}

%\section*{Acknowledgments}
%Include acknowledgments of funding, any patents pending, where raw data for the paper are deposited, etc.
   
%\section*{Supplementary materials}
\section*{Materials and Methods}
\subsection*{Preprocessing of electrocorticogram data}
Multi-day electrocorticogram recordings from 17 patients undergoing presurgical monitoring at the Epilepsy Center of the University Hospital of Freiburg, Germany \cite{Ihle2012} were analyzed. The number and dosing level of antiepileptic drugs (AEDs) varied over the course of the recording period. Patients gave informed consent. The number of electrodes varied between patients and included both subdural and depth electrodes (mean number of electrodes $n=78\pm24$, 30-121). Electrode placement was solely determined by clinical considerations. Electrocorticogram data were sampled at either 256 Hz, 512 Hz or 1024 Hz. If sampled at higher rate, data were first downsampled to 256 Hz. A notch filter was then applied to remove potential contamination with 50 Hz line noise. Data was preprocessed in segments of one hour duration. To compare high and low AED medication regimes, we picked the one day with the highest cumulative AED load and the one day with the lowest AED load in each patient and analyzed all hours from midnight to midnight within this day. If there were more than one day with highest and/or lowest medications, we picked the two days furthest apart from each other.

\subsection*{Detection of epileptic spikes}
Spikes are large, abnormal discharges that occur between seizures in patients with epilepsy. We here detected sharply contoured waveforms as spikes via a previously validated method (Fig. \ref{fig_2} A, \cite{Barkmeier2012}). In brief, for each one-minute long data block, potential spikes were detected, if they cross a threshold defined by standard deviations (SD coefficient=4) of the absolute amplitude of high bandpass filtered signal (20–50 Hz) for the channel. 
Next, the raw ECoG data were bandpass filtered between 1–35 Hz (second order digital Butterworth) and all channels in the one-minute block were scaled by a  scaling factor which is the median value of the average absolute amplitudes across all channels in the grid.
Once data had been scaled, shape criteria of amplitude, duration, and slope were applied to the scaled, lower bandpass filtered EEG signal (1–35 Hz) at the previously identified potential spikes. 
Spike duration was determined by searching 10 sampling steps (at 256 Hz) on either side of the detected peak to find the minima on each side. Standard parameters described previously were used to identify spikes \cite{Barkmeier2012}: total amplitude of both half-waves $>600 \mu V$, slope of each half-wave $>7 \mu V/ms$, duration of each half-wave $>10 ms$. Timestamps of spikes were saved for further analysis.

\subsection*{Detection of spike cascades}
Fig. \ref{fig_2} B shows the distribution of inter-spike intervals from one patient. The bimodality indicates a fast timescale belonging to the inter-spike intervals within a cascade (left peak), and a much slower timescale indicating the intervals between cascades (right peak). A cascade was defined as a spatiotemporal cluster of consecutive spikes with inter-spike intervals not exceeding a temporal threshold $􏰀\Delta T$. $\Delta T$ was chosen to be in the trough between the two peaks $\Delta T = 20$ sampling steps at 256 Hz (grey vertical line) in order to identify spikes belonging to one cascade and to prevent concatenation of separate cascades.

\subsection*{Signal autocorrelation}
Modulations in signal power are generally useful to characterize neural dynamics \cite{Donner2011}. In particular, fluctuations in the broadband high-frequency 50–200 Hz range have been shown to provide a local, spatiotemporal estimate of population spike rate variations near an electrode \cite{Manning2009, Miller2010, Nir2007, Ray2011, Whittingstall2009}.

For each ECoG channel, time-courses of broadband high-frequency power fluctuations were obtained by computing the mean 50-100 Hz signal power every 125 ms (FFT routine, Hanning window) for each hour during either a high or low AED medication day. Signal power estimates are not normally distributed across time samples, the logarithm of power estimates was thus taken to normalize their distributions \cite{Miller2009,Honey2012}. An autocorrelation function of the power time course was then obtained for each electrode and hour in recording. Analyses in the main part of the manuscript are based on average autocorrelation functions across all electrodes and across all hours in either high or low AED medication day.

ACW was defined as the full-width-at-half-maximum of the autocorrelation function of the power time course. For this purpose, autocorrelation functions across all channels and hours in either high/low medication day were averaged for each patient. ACW was determined as twice the time lag at which the ACF became smaller than half its value between maximum to minimum. Since time lags are in 125 ms increments, the minimal value of ACW is 250 ms. 

\subsection*{Computational neuron network model}
The neuron network model consists of N=200 binary-state neurons connected by all-to-all, asymmetric synaptic coupling strengths $w_{ij}$. Each neuron j is either excitatory or inhibitory, respectively corresponding to $w_{ij}\ge0$ or $w_{ij}\le0$ for all i. 
The binary state $s_i(t + 1)$ of neuron i (s = 0 inactive, s = 1 spiking) is determined based on the sum $p(t + 1)$ of its inputs $p(t + 1) = \sum_{j=1}^{N} w_{ij}(t)s_j(t)$ and the parameter controlling neuronal excitability, $p_{ne}$, according to the following dynamical rules: if $0<p*p_{ne}<1$, then the neuron fires with probability $p*p_{ne}$, if $p*p_{ne}\ge 1$, then the neuron fires with probability 1, if $p*p_{ne}\le 0$, then the neuron does not fire. 

$w_{ij}$ values are first drawn from a uniform distribution [0,1]. Then, 20\% of neurons are set to being inhibitory (by multiplication of the corresponding coupling strengths with -1) and the remaining 80\% of neurons to being excitatory. $p_{ij}$ are then multiplied by $N \cdot K /\sum w_{ij}$. The parameter $K$ consequently captures the average connectivity and is closely related to the largest eigenvalue $\lambda$ of the adjacency matrix $w_{ij}$ which controls the dynamics of the network: at $K=1$ each spiking neuron excites, on average, exactly one postsynaptic neuron meaning the network is critical \cite{Larremore2014,Meisel2017b}. Conversely, at $K<1$ activity dies out prematurely and the system is subcritical; at $K>1$ each neuron excites on average more than one postsynaptic neuron and the system is supercritical. 

\begin{figure}%[htbp]
\centering
\includegraphics[width=0.65\textwidth]{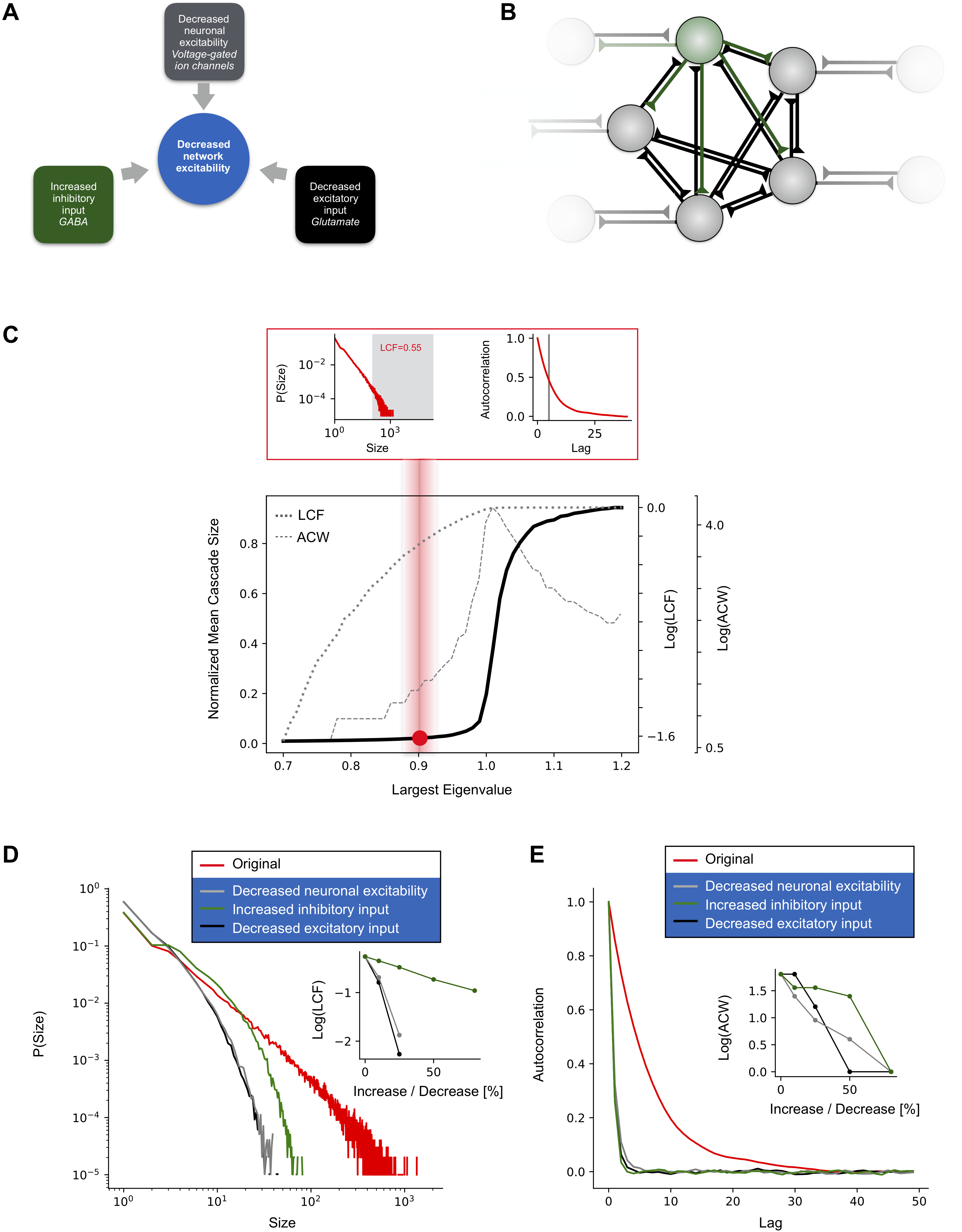}
\caption{Antiepileptic drug (AED) action induces subcritical dynamics in a neural network model. A, Illustration of the main mechanisms of AED action. Collectively, AEDs aim to reduce seizure risk by decreasing cortical network excitability. B, Conceptual cartoon illustrating neural network model features, including excitatory (black) and inhibitory (green) recurrent synapses. The strengths of inhibitory and excitatory synaptic transmission along with neuron excitability can be selectively changed to mimic AED action. C, Network dynamics exhibits a phase transition between an inactive phase, where cascades remain small and local and an active phase, where activity is dominated by large cascades spanning the whole network upon increasing connection strengths (solid black line). Grey dashed line indicates the large cascade fraction, LCF. Grey dotted line indicates autocorrelation function half-width, ACW, which peaks at criticality ($\lambda=1$). Red area inset shows cascade size distribution and autocorrelation functions when dynamics is poised in a slightly subcritical regime, mimicking experimental observations. D, E, AED action incurs decline of large cascade sizes and faster autocorrelation function decline. Large plots show exemplary cascade size distribution and autocorrelation function (50\% decrease in neuron excitability, 50\% decrease in excitatory synaptic strength, ten fold increase in inhibitory synaptic strength). Insets show LCF and ACW for a range of network excitability reducing parameter values.\label{fig_1}
}
\end{figure}

The model affords implementation of different AED mechanisms known to reduce cortical network excitability: reduction of individual neuronal excitability, reduction of excitatory synaptic transmission and increase of inhibitory synaptic transmission (Fig. \ref{fig_1} A, \cite{Bialer2010}). Reduction of individual neuronal excitability is controlled by parameter $p_{ne}$, where $p_{ne}\le 1$. Decreased excitatory synaptic transmission is modelled by multiplication of all positive $w_{ij}$ with factor $p_{exc}$, where $p_{exc}\le 1$. Increased inhibitory synaptic transmission is modelled by multiplication of all negative $w_{ij}$ with factor $p_{inh}$, where $p_{inh}\ge 1$. 

The onset of stimulation is instantiated by setting a random neuron to active. Activity is monitored until no neuron is active anymore or, in case of ongoing activity, until 500 time steps have passed, at which point a new cascade is started by setting a random neuron to active. We modelled a total of 100000 such cascades at each connectivity $K$. The temporal autocorrelation was studied using the time course of overall network activity, i.e., the sum of active neurons at each time step.

\clearpage

\newpage

\newpage

\newpage

\newpage
%\noindent {\bf Fig. 1.} Please do not use figure environments to set
%up your figures in the final (post-peer-review) draft, do not include graphics in your
%source code, and do not cite figures in the text using \LaTeX\
%\verb+\ref+ commands.  Instead, simply refer to the figure numbers in
%the text per {\it Science\/} style, and include the list of captions at
%the end of the document, coded as ordinary paragraphs as shown in the
%\texttt{scifile.tex} template file.  Your actual figure files should
%be submitted separately.

\end{document}